\documentclass[aps,prl,superscriptaddress,twocolumn,nofootinbib,preprintnumbers]{revtex4}

\usepackage{amsmath}
\usepackage{graphicx,tabularx}
 
\newcommand{\sla}[1]{/\!\!\!\!#1}
 
\newcommand{\ttt}[1]{\texttt{#1}}
\newcommand{\mrm}[1]{\ensuremath{\mathrm{#1}}}
\newcommand{\tsc}[1]{\textsc{#1}}
 
\renewcommand{\t}{t}
\newcommand{\tbar}{\bar{t}}
\newcommand{\qbar}{\bar{q}}

\newcommand{\go}   {\tilde{g}}

\newcommand{\sq}[1]{\tilde{q}_{{#1}}}
\newcommand{\suL}{\tilde{u}_{{L}}}
\newcommand{\se}[1]{\tilde{\ell}_{{#1}}}
\newcommand{\nn}[1]{\tilde{\chi}^0_{{#1}}}
 
\providecommand{\SP}{\scriptscriptstyle}
\newcommand{\mgo}   {m_{\tilde{g}}}

\newcommand{\msu}[1]{m_{\tilde{u}_{\SP {#1}}}}

\newcommand{\pT}{\ensuremath{p_T}}
\newcommand{\pTs}{\ensuremath{p_T^2}}
\newcommand{\pTj}{\ensuremath{p_{T,j}}}
 
\def\ie{{\it i.e.\;}}
\def\eg{{\it e.g.\;}}

\newcommand{\gev}{~{\ensuremath\rm GeV}}

\newcommand{\pb}{~{\ensuremath\rm pb}}

\begin{document}
 
\date{\today}
 
\title{Squark and Gluino Production with Jets}

\preprint{MPP-2005-101, FERMILAB-PUB-05-352-T}
 
\author{T.~Plehn}
\affiliation{Heisenberg Fellow, Max Planck Institute for Physics, Munich, Germany}
\author{D.~Rainwater}
\affiliation{Marshak Fellow, Dept.of Physics and Astronomy, University of Rochester,
             Rochester, USA}
\author{P.~Skands}
\affiliation{Theoretical Physics Dept., Fermi National Accelerator Laboratory,
             Batavia, USA}
 
\begin{abstract}
  We present cross section predictions for squark and gluino
  production at the LHC, in association with up to two additional hard
  jets.  These cross sections can be very large in comparison to the
  inclusive Born rates.  Because hadron collider experiments utilize
  hard jets in the reconstruction of cascade decays or as a way to
  separate squark and gluino production, the understanding of these
  processes is crucial.  We show to what degree hard jet radiation can
  be described by shower algorithms and point out how tuning these
  showers, for example to top quark pair production, could help reduce
  theoretical uncertainties for new physics searches at the LHC.
\end{abstract}

\maketitle
 

In the Standard Model (SM), the mechanism of the observed electroweak
symmetry breaking is widely believed to involve a Higgs boson. This
fundamental scalar poses a theoretical problem: the stability of its
mass after the inclusion of radiative corrections.  A possible
solution is TeV-scale supersymmetry.  The minimal supersymmetric
extension of the Standard Model (MSSM) simultaneously solves several
problems in high energy physics and cosmology: gauge coupling
unification; radiative electroweak symmetry breaking~\cite{susy}; and
a stable weakly-interacting dark matter candidate~\cite{dark_matter}.

With the Tevatron in operation and the LHC only a few years distant,
the TeV scale is rapidly coming within reach.  Squark and gluino
production cross sections approach ${\cal O}(\rm{pb})$ at the Tevatron
and ${\cal O}(\rm{nb})$ at the LHC, for masses around the present
Tevatron exclusion limits of up to $400\gev$.

\smallskip
 
\underline{MSSM searches and jets:} The main difference between 
R-parity-conserving MSSM signals and SM QCD backgrounds at a hadron
collider comes from the stable lightest supersymmetric particle, an
end product of all superpartner decays, which escapes the detector
unobserved.  Requiring a large amount of missing transverse energy is
thus the first ingredient to enhance the signal.  The QCD-strength
production channels for squarks and gluinos are $pp\to\go\go,
\sq{}\sq{}^*, \sq{}\sq{}, \sq{}\go$~\cite{deq,nlo}.  To a good
approximation, the light-flavor MSSM squarks are mass degenerate,
while the lighter of the two top squarks is often the lightest
strongly-interacting MSSM particle.  Searching for squarks and gluinos
in an inclusive analysis the signature is jets plus $\sla{E}_T$,
possibly plus leptons.  The shortest cascade of two-particle decays is
$\go\to\sq{}\bar{q}$ and $\sq{}\to q\nn{1}$, where the lightest
neutralino $\nn{1}$ is stable.  Such an inclusive search is
well-suited to find MSSM-type deviations from the Standard Model.
Because the gluino decay gives one more hard jet in the final state,
an event's jet multiplicity provides discrimination between the
relative rates of squarks and gluinos.
 
\begin{table}[b]
\begin{tabular}{|r|c|rrrrr|}
 \hline
  & $\sigma_{\rm tot} [\rm{pb}]$
  & $\tilde{g} \tilde{g}$
  & $\tilde{u}_L \tilde{g}$
  & $\tilde{u}_L \tilde{u}_L^*$
  & $\tilde{u}_L \tilde{u}_L$
  & $T\bar{T}$ \\
  \hline \hline
  $\pTj>100\gev$ &$\sigma_{0j}$& \;4.83 & \;5.65 & \;0.286 & \;0.502 & \;1.30\\
                 &$\sigma_{1j}$& 2.89 & 2.74 & 0.136 & 0.145       & 0.73\\
                 &$\sigma_{2j}$& 1.09 & 0.85 & 0.049 & 0.039 & 0.26\\
  \hline \hline
  $\pTj>50\gev$  &$\sigma_{0j}$& 4.83 & 5.65 & 0.286 & 0.502 & 1.30\\
                 &$\sigma_{1j}$& 5.90 & 5.37 & 0.283 & 0.285 & 1.50\\
                 &$\sigma_{2j}$& 4.17 & 3.18 & 0.179 & 0.117 & 1.21\\
  \hline
\end{tabular}
\caption{Cross sections for the production of a toy-model $600\gev$
top quark, squarks and gluinos at the LHC
at the benchmark point SPS1a. We show fixed-order matrix
element results with 0,1,2 additional hard jets, and with two
different $p^{\rm min}_{T,j}$ values and $|y_j|<5$ and $R_{jj}>0.4$.}
\label{tab:njet}
\end{table}
 
We can also make use of longer decay chains, \eg of the classic type
$\go\to\sq{}\bar{q}\to\nn{2}q\bar{q}\to\se{}\bar{\ell}q\bar{q}\to\nn{1}\ell\bar{\ell}q\bar{q}$
with five unknown masses.  These masses can be extracted from
kinematic distributions, \ie thresholds and edges of different
momentum combinations~\cite{edges}.  Alternative methods have been
developed to improve the mass reconstruction~\cite{nojiri} and the
associated statistical and systematic errors. These measurements can
in turn be used to determine parameters of the TeV-scale MSSM
Lagrangian~\cite{sfitter}.

\smallskip
 
To make optimal use of the achievable statistical precision, as well
as to quantify the errors on extracted model parameters, it is crucial
to properly understand the systematic errors in the cascade
reconstruction.  Obviously, effects such as detector resolution and
jet energy scale calibration will have a large impact~\cite{david}.
In this letter, we focus on another source of (combinatorial) error:
the presence of additional, observable hard jets due to SM QCD
radiation, with transverse momenta comparable to the typical cuts
planned for squark and gluino studies, about
$p_T>50-100$~GeV~\cite{ATLAS,CMS}.  Extra jets, in particular from the
initial state, will introduce noise when reconstructing the cascade
kinematics from the observed jet kinematics and when attempting to
separate squark- and gluino-enriched samples.  This question has a
counterpart in the hadronic top analyses at the Tevatron.  We know
from data that in these analyses additional jet radiation fakes $W$
decay jets in a non-trivial fraction of the events~\cite{datajets}.
With generally larger energy scales involved in SUSY processes, one
would expect the initial state to be more active, hence potentially
more dangerous.
\medskip
 
\begin{figure*}[t]
\begin{center}
\scalebox{0.9}{
\begin{minipage}{\textwidth}
\begin{center}
\vspace*{-12mm}
\hspace*{-5mm}\includegraphics[scale=0.67]{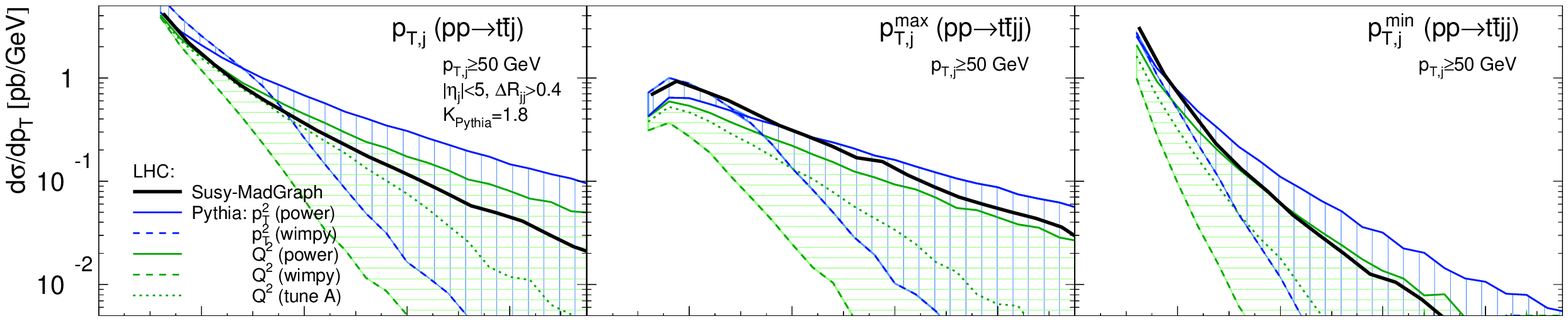}\hspace*{-5mm}
\vspace*{-16mm}
\vspace*{-11mm} 
\hspace*{-5mm}\includegraphics[scale=0.67]{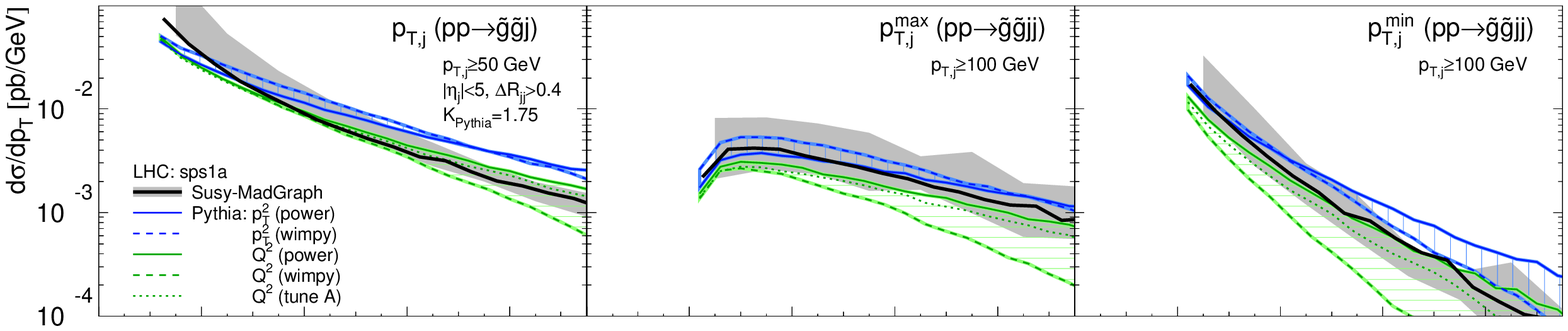}\hspace*{-5mm}
\vspace*{-16mm}
\vspace*{-11mm}
\hspace*{-5mm}\includegraphics[scale=0.67]{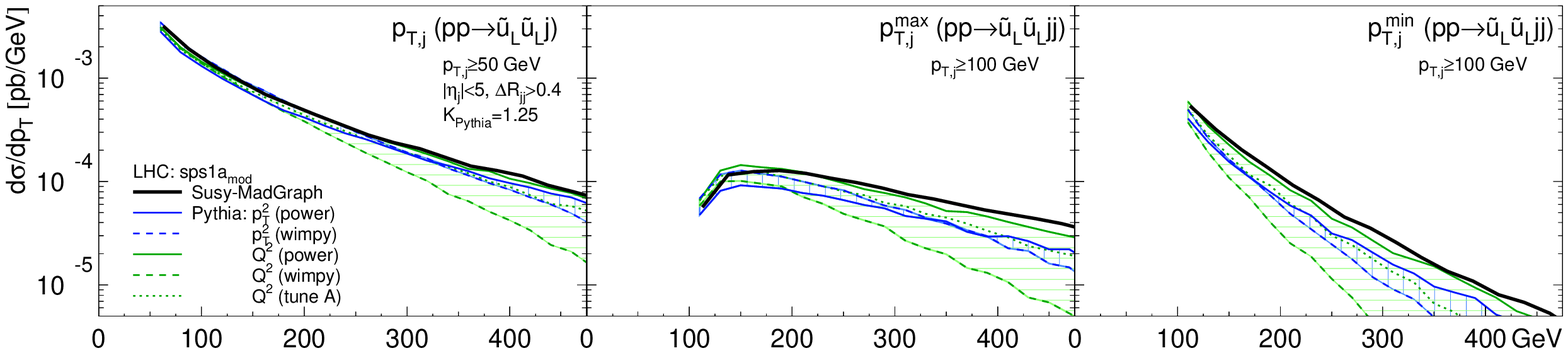}\hspace*{-5mm}
\vspace*{-12mm}
\end{center}
\end{minipage}}
\end{center}
\caption{$\pTj$ spectra for $\t\tbar,\go\go$ and $\suL\suL$ production
  in association with 1 and 2 hard jets.  NLO K-factors~\cite{nlo} are
  applied to the \textsc{Pythia} results to avoid a normalization
  mismatch in the one-jet case.  The mass spectrum is given by SPS1a,
  except for $\suL\suL$ where we reduced the gluino mass, as in
  Tab.~\ref{tab:njet}. At high $\pT$ (of order the factorization scale
  and above), the matrix elements are, by definition, the most
  reliable while the parton showers can be seen to be associated with
  large uncertainties, whereas the opposite is the case at low $\pT$.
  The shaded region in the middle row is the theoretical uncertainty
  on the matrix element prediction.  See text for details.}
\label{fig:compare}
\vspace{-4mm}
\end{figure*}

\underline{Hard jets from matrix elements:} To study the production of
hard jets with squarks and gluinos at the LHC, we first present the
results of a fixed-order approach.  We use the new supersymmetric
version~\cite{smg,catpee} of the event generator
MadEvent~\cite{madevent} to calculate tree-level rates for $pp \to
\go\go, \sq{}\sq{}^*, \sq{}\sq{}, \sq{}\go$, including the emission of
one and two additional jets.  To avoid regions sensitive to the soft
and collinear singularities of initial- and final-state radiation we
limit ourselves to $\pTj>50\gev$ (and in some cases $\pTj>100\gev$),
for which we expect fixed-order perturbation theory to be reliable at
the LHC.

\smallskip
 
In Table~\ref{tab:njet} we show inclusive production cross section
estimates for squarks and gluinos plus zero to two hard jets, for the
parameter point SPS1a, where $\mgo=608\gev$ and
$\msu{L}=567\gev$~\cite{sps}.  The factorization scale is set to the
average final state mass, as is the renormalization scale for the
heavy pair.  The renormalization scale for additional jet radiation is
$\pTj$, the standard choice in shower Monte Carlos~\cite{ptscale}.  We
see that for jets with $p_T>100\gev$ the perturbative expansion is
stable, but the relative suppression is closer to $1/2\cdots 1/3$ than
to $\alpha_s/\pi$.  If we allow semi-hard jets down to $50\gev$ the
fixed-order perturbative expansion approaches its limit.  We
emphasize that it is not a problem for the one-jet rate to be
slightly larger than the all-inclusive rate at leading order, as long
as it is in the range of the NLO inclusive cross section~\cite{nlo}.
To check that this behavior is quantitatively universal for heavy QCD
production we also show the rates for a heavy toy-quark $T$
(effectively a 600\gev\ top), and we indeed see the same
pattern.

\smallskip

The case of gluino pair production exemplifies that gluon radiation
(mostly from the initial state) is indeed the dominant source of extra
jets: allowing just initial- and final-state gluons with
$\pTj>50\gev$, the $\go\go$ case in Table~\ref{tab:njet} becomes
$\sigma_{0,1,2 j} = (4.63, 3.90, 2.03)~\pb$.  In addition, jet
radiation probes new initial states through crossing of quarks and
gluons from the final state to the initial state: \eg purely
quark-initiated processes like $qq\to\sq{}\sq{}$ receive large
corrections from $qg$ scattering.  These crossed channels include
intermediate states of the kind $qg\to\sq{}\go\to\sq{}\sq{}\qbar$.
Usually these intermediate states are subtracted in the narrow-width
approximation to avoid double counting between NLO $\sq{}\sq{}$ and
$\go\sq{}$ production~\cite{nlo}.  Here, we explicitly remove all
on-shell intermediate gluinos from $\go\suL$ production.  For the
$\suL\suL^{(*)}$ channel we simply decrease the gluino mass to
$558\gev$ to avoid intermediate on-shell gluinos.

\medskip
 
\underline{Jets from parton showers:} To compare to experimental
observables, multiple soft emission and hadronization effects can be
important -- aspects which go beyond the scope of a fixed-order
calculation.  For bremsstrahlung emission, logarithms
$\alpha_s^N\log^{2N}(Q^2_{\mrm{soft}}/Q^2_{\mrm{hard}})$ appear to all
orders in perturbation theory, while the corrections associated with
hadronization are inherently non-perturbative, $\alpha_s \to 1$.
Exploiting collinear factorization in QCD and assuming universality of
hadronization, both types of corrections may be included, at least
approximately.  This is the basis of multi-purpose event generators
like \tsc{Herwig}, \tsc{Pythia}, or \tsc{Sherpa}.  Starting from a
given set of final state partons, a sequence of initial- and
final-state QCD branchings are generated, resumming the leading
logarithms mentioned above. The emissions are ordered, \eg in
the parton virtuality $Q$ or in $\pTj$, and the description is
matched to hadronization models at a fixed low scale, $\sim
1\gev$.

Especially for hard radiation, large differences may exist between
different shower algorithms.  To quantify, we use
\tsc{Pythia}~\cite{pythia} with two qualitatively different showers:
one $Q^2$-ordered~\cite{q2showers}, and the other 
$\pT$-ordered~\cite{Sjostrand:2004ef}.  Note that, due to the
large final state squark and gluino masses, we explore mainly the
properties of the initial-state showers.  The crucial parameter
here is the starting scale of the shower, which sets an upper limit to
the phase space over which jets can be radiated.  For initial-state
radiation, this starting scale is nominally identical to the
factorization scale, where the parton densities are convoluted with
the matrix elements.  In \tsc{Pythia}, this scale is normally the
transverse mass $\mu_F=\sqrt{\pT^2+ \hat{m}^2}$, with $\hat{m}$ the
average mass of the final state SUSY particles, and $\pT$ their
relative transverse momentum.

For a $\pT$-ordered shower, this $\mu_F$ can be used directly as the
maximum $\pTj$.  We refer to this choice as the $\pT$-ordered `wimpy
shower'.  We also investigate the consequences of allowing the parton
shower to populate the full phase space, with the maximum
$\pTj=\sqrt{s}/2$, regardless of $\mu_F$.  This choice we refer to as
the $\pT$-ordered `power shower'.  Though strictly-speaking in
conflict with the factorization assumption, this choice has
interesting phenomenological consequences, as we shall see.

The case of a $Q^2$-ordered shower is not so simple.  The starting
scale here is $Q^2_{\mrm{max}} = \min\left(C \mu_F^2, s\right)$, where
$C\ge 1$ parameterizes the translation from $\pTs$ to $Q^2$.  We refer
to $C = 1$ as the $Q^2$-ordered wimpy shower, $C=4$ as
Tune~A~\cite{tunea} (designed to match Tevatron data), and
$C\to\infty$ as power shower --- with the same caveat concerning
factorization as for the $\pT$-ordered version.

\medskip
 
\underline{Numerical Comparison:} We now turn to a comparison of the
three processes $\t\tbar$+jets, $\go\go$+jets, and $\suL\suL$+jets, at
the LHC. Here, $\t\tbar$ represents a process with a 
small ratio $\mu_F/\sqrt{s}$, while the SUSY processes involve much
larger masses and hence larger factorization scales.  The difference
between $\go\go$ and $\suL\suL$ is the sea- v.\ valence-dominated
initial state.

Jets in the parton shower final states were clustered using a cone
algorithm with $\Delta R=0.4$, \ie\ similar to the ME cut $R_{j,j}>0.4$. 
By looking at $p_{T,j}/p_{T,\mathrm{ME}}$ in generated dijet samples,
we checked explicitly that out-of-cone corrections are too small to
affect our study significantly. 

In Fig.~\ref{fig:compare} we show results for the $\pTj$ spectra,
using CTEQ5L parton distribution functions~\cite{Lai:1999wy}.  We use
the heavy particle mass as the central value for the factorization
scale.  For renormalization scale, we apply one factor of $\alpha_s$
using $Q^2=m_T^2$ for each final-state particle.  The shaded band in
the middle row highlights the theoretical uncertainty by plotting the
extremal values of four factorization and renormalization scale
choices: varying $\mu_R$ down to the minimum $p_T$ of the jets, and
varying $\mu_F$ down to the minimum $p_T$ of the jets and up to
$\sqrt{\hat{s}}/2$.

Consider first $\t\tbar$+1jet (upper left): as a general feature, the
power showers exhibit a harder high-$\pT$ spectrum than the matrix
element.  In contrast, note the cataclysmic drop of the wimpy showers
above $\pTj\sim m_t\sim \mu_F$, with the $Q^2$-ordered wimpy showers
everywhere softer than the matrix element.  Finally, Tune A is
indistinguishable from the matrix element for $\pTj < 250\gev$, this
region being similar to the phase space accessible at the Tevatron.
All the calculations agree fairly well down to $\pTj\sim 50\gev$, a
value sufficiently below the factorization scale for the parton
showers to be reliable, but still large enough for stable fixed-order
predictions.  The crossover at $\sim150\gev$ between the two
$\pT$-ordered showers illustrates the effect of changing $\mu_R$ from
$\pT/2$ in the wimpy shower to $3\pT$ in the power shower.  The
$Q^2$-ordered showers all use $\mu_R=\pTj$.
  
For the two-jet distributions, note that the collinear approximation
is only rigorously correct in the limit that each successive jet is
much softer than the preceding one, hence the two-jet shower
predictions are associated with even larger uncertainties.
Nevertheless, the power showers deviate from the matrix element by
less than a factor two over most of the $\pTj$ range.  Although
remaining differences could undoubtedly be cured by slight shower
parameter modifications, we note that obtaining simultaneous agreement
of the one- and two-jet rates is likely to require more sophisticated
matching techniques~\cite{matching,mrw,mcnlo,sherpa}.

\smallskip

The parton shower predictions for the SUSY processes exhibit
significantly less variation, as shown in the lower rows of
Fig.~\ref{fig:compare}.  Owing to much larger factorization scales the
presence or absence of a cutoff in the parton shower evolution at
$\mu_F$ does not lead to very large differences for the $\pTj$ regions
we consider.  In this way, the kinematic regime of squark and gluino
production at the LHC is more similar to \eg\ $t\tbar$ production at
the Tevatron than at the LHC.  Considering first $\go\go$+1~jet, we
observe that the matrix element rate begins to diverge around $\pTj
\sim 100\gev$, due to large logarithms, $\log^2 m^2_{\go}/\pTj^2$.
For $Z$, light Higgs, or even $t\tbar$ production this breakdown scale
is much smaller, of order the minimum $\pTj$ observable at the
LHC~\cite{resummation}.  The case of heavy particle production, such
as gluinos and squarks, is different: for jets softer than
$\pT\sim100\gev$ we would be well-advised to include
resummation effects.
 
At large values of $\pTj$, a pattern similar to the tops arises: the
one-jet radiation is very well described by Tune A. For two-jet
radiation, the $Q^2$-ordered showers generally fall below the matrix
element, while the $\pT$-ordered ones overshoot.  We also studied
$\suL \suL^*$ and $\suL \go$ production and found that they exhibit
essentially the same behavior as gluino-pair production.

For the last process in Fig.~\ref{fig:compare}, $\suL\suL$ production,
the matrix element divergence at low $\pT$ appears to be much milder
than for the $\go\go$ case.  We interpret this as a consequence of the
less radiating, valence-dominated initial state.  For the high-$\pT$
tail, the power showers are again in fairly good agreement with the
matrix element, though with a much smaller difference between $\pT$-
and $Q^2$-ordered showers than above.

\begin{figure}[t]
\vspace*{-10mm}\includegraphics[scale=0.60]{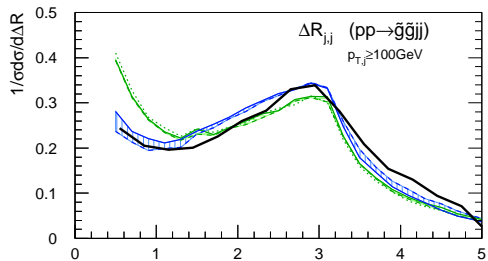}
\vspace*{-12mm}
\caption{The $\Delta R_{jj}$ distribution for $\go\go jj$ production
  as predicted by the hard matrix element and the parton showers.}
\label{fig:angular}
\vspace{-4mm}
\end{figure}

Lastly, we compare the $\Delta R_{jj}$ distributions for gluino pairs
plus two jets in Fig.~\ref{fig:angular}.  The results are not
drastically different. The difference between the two different shower
models at low $\Delta R$ is interesting, however. It should probably
not be interpreted as the onset of a collinear singularity. In that
case, one would expect the showers to agree with each other, but not
with the matrix elements. Moreover, from the results presented in
Fig.~\ref{fig:compare} it is clear that with a jet cut of 100\gev\ we
are nowhere near the collinear region. In fact, the cutoff itself may
furnish part of the reason; the region where both jets are close to
the cutoff is, by definition, not strongly ordered, and would hence be
expected to be problematic for shower descriptions.  We plan to return
to this in a future study.

\medskip
 
\underline{Summary:} Using SUSY--MadEvent we show that
matrix--element-based QCD calculations predict a large number of hard
jets associated with squark and gluino production at the LHC.  This
effect should be taken into account in studies of the separation of
squark and gluino event samples, and for cascade decay reconstruction.
We compared in detail the matrix element approach and the results from
$\pT$-ordered and $Q^2$-ordered showers, implemented in
\tsc{Pythia}~6.3.

For the radiation of one extra jet in SUSY heavy colored pair
production processes, conventional parton showers with a phase space
cutoff at the factorization scale give a reasonable approximation up
to $\pTj \sim \mu_F/2$, above which they rapidly break down.  This is
similar to what has been observed before for
Drell--Yan~\cite{mrw,wplusjets,q2merge} (and Higgs) production.  A
significant improvement can be obtained by removing the explicit phase
space cut, allowing the shower to populate all of phase space.
However, this tends to yield somewhat harder radiation spectra than
produced by the matrix elements, again akin to what has been shown for
hadroproduction of colorless resonances~\cite{q2merge}.

For fairly soft jets, we see that in the production of high-mass
gluinos the breakdown of fixed-order perturbation theory caused by
logarithmic corrections can occur already at jet transverse momenta of
as high as 100~GeV.

\bigskip

 
\begin{acknowledgments}
  We would like to thank Kaoru Hagiwara and Tim Stelzer for their
  great help creating SUSY-MadEvent; Torbjorn Sj\"ostrand for
  enlightening discussions and comments on the manuscript; and finally
  the Madison Pheno group, the DESY and KEK theory groups, and the
  Aspen Center for Physics for their generous hospitality.  This
  research was supported in part by the U.S. Department of Energy
  under grant Nos. DE-FG02-91ER40685 and DE-AC02-76CH03000.
\end{acknowledgments}
 


\baselineskip15pt
 
\begin{thebibliography}{99}
 
\bibitem{susy}
 For reviews of SUSY, see \eg:
 I.~J.~R.~Aitchison,
  hep-ph/0505105;
 S.~P.~Martin,
  hep-ph/9709356.

\bibitem{dark_matter}
 J.~R.~Ellis, J.~S.~Hagelin, D.~V.~Nanopoulos, K.~A.~Olive and M.~Srednicki,
  Nucl.\ Phys.\ B {\bf 238}, 453 (1984);
 H.~Goldberg,
  Phys.\ Rev.\ Lett.\  {\bf 50}, 1419 (1983);
 for a recent review see \eg:
 G.~Bertone, D.~Hooper and J.~Silk,
  Phys.\ Rept.\  {\bf 405}, 279 (2005).


\bibitem{deq}
 S.~Dawson, E.~Eichten and C.~Quigg,
  Phys.\ Rev.\ D {\bf 31}, 1581 (1985).
 
\bibitem{nlo}
 W.~Beenakker, R.~H\"opker, M.~Spira and P.~M.~Zerwas,
  Nucl.\ Phys.\ B {\bf 492}, 51 (1997);
 W.~Beenakker, M.~Kr\"amer, T.~Plehn, M.~Spira and P.~M.~Zerwas,
  Nucl.\ Phys.\ B {\bf 515}, 3 (1998);
 W.~Beenakker, M.~Klasen, M.~Kr\"amer, T.~Plehn, M.~Spira and P.~M.~Zerwas,
  Phys.\ Rev.\ Lett.\  {\bf 83}, 3780 (1999).
 
\bibitem{edges}
 H.~Bachacou, I.~Hinchliffe and F.~E.~Paige,
  Phys.\ Rev.\ D {\bf 62}, 015009 (2000);
 B.~C.~Allanach, C.~G.~Lester, M.~A.~Parker and B.~R.~Webber,
  JHEP {\bf 0009}, 004 (2000).
 
\bibitem{nojiri}
 K.~Kawagoe, M.~M.~Nojiri and G.~Polesello,
  Phys.\ Rev.\ D {\bf 71}, 035008 (2005).
 
\bibitem{sfitter}
 R.~Lafaye, T.~Plehn and D.~Zerwas,
  hep-ph/0404282;
 P.~Bechtle, K.~Desch and P.~Wienemann,
  hep-ph/0412012.
 
\bibitem{david}
 B.~K.~Gjelsten, D.~J.~Miller and P.~Osland,
  JHEP {\bf 0412}, 003 (2004)
 and
  hep-ph/0501033.

\bibitem{ATLAS}
  ATLAS TDR, report CERN/LHCC/1999-15 (1999).

\bibitem{CMS}
  CMS TDR, report CERN/LHCC/2006-001 (2006).

\bibitem{datajets}
  L.~Orr, private communication; R.~Demina, private communication.

\bibitem{smg}
  G.-C.~Cho, K.~Hagiwara, J.~Kanzaki, T.~Plehn, D.~Rainwater and T.~Stelzer,
  in preparation.
 
\bibitem{catpee}
  K.~Hagiwara, W.~Kilian, F.~Krauss, T.~Ohl, T.~Plehn, D.~Rainwater J.~Reuter 
  and S.~Schumann, hep-ph/0512260.
 
\bibitem{madevent}
  T.~Stelzer, F.~Long,
  Comput.{} Phys.{} Commun.{} \textbf{81} (1994) 357;
  F.~Maltoni and T.~Stelzer,
  JHEP {\bf 0302}, 027 (2003).
 
\bibitem{sps}
 B.~C.~Allanach {\it et al.},
  Eur.\ Phys.\ J.\ C {\bf 25}, 113 (2002).
 
\bibitem{ptscale}
 D.~Amati, A.~Bassetto, M.~Ciafaloni, G.~Marchesini and G.~Veneziano,
  Nucl.\ Phys.\ B {\bf 173}, 429 (1980);
 G.~Curci, W.~Furmanski and R.~Petronzio, 
  Nucl.\ Phys.\ B {\bf 175}, 27 (1980).

\bibitem{pythia}
  T.~Sj\"ostrand, P.~Eden, C.~Friberg, L.~L\"onnblad, G.~Miu, S.~Mrenna and E.~Norrbin,
  Comput.\ Phys.\ Commun.\  {\bf 135}, 238 (2001)
 T.~Sj\"ostrand, L.~L\"onnblad, S.~Mrenna and P.~Skands,
  hep-ph/0308153.
 
\bibitem{q2showers}
 T.~Sj\"ostrand,
  Phys.\ Lett.\ B {\bf 157}, 321 (1985);
  M.~Bengtsson, T.~Sj\"ostrand and M.~van Zijl,
  Z.\ Phys.\ C {\bf 32}, 67 (1986);
  M.~Bengtsson and T.~Sj\"ostrand,
  Phys.\ Lett.\ B {\bf 185}, 435 (1987)
 and
  Nucl.\ Phys.\ B {\bf 289}, 810 (1987).

\bibitem{Sjostrand:2004ef}
 T.~Sj{\"o}strand and P.~Z.~Skands,
  Eur. Phys. J. {\bf C39}, 129-154 (2005).

\bibitem{tunea}
 R.D. Field,
  hep-ph/0201192
  CDF Note 6403;
 further recent talks available from webpage
  \ttt{http://www.phys.ufl.edu/}$\sim$\ttt{rfield/cdf/}
 
\bibitem{Lai:1999wy}
  H.~L.~Lai {\it et al.}  [CTEQ Collaboration],
  Eur.\ Phys.\ J.\ C {\bf 12}, 375 (2000).

\bibitem{tops}
 P.~Nason, S.~Dawson and R.~K.~Ellis,
  Nucl.\ Phys.\ B {\bf 303}, 607 (1988);
 M.~L.~Mangano, P.~Nason and G.~Ridolfi,
  Nucl.\ Phys.\ B {\bf 373}, 295 (1992);
 W.~Beenakker, W.~L.~van Neerven, R.~Meng, G.~A.~Schuler and J.~Smith,
  Nucl.\ Phys.\ B {\bf 351}, 507 (1991);
 R.~Bonciani, S.~Catani, M.~L.~Mangano and P.~Nason,
  Nucl.\ Phys.\ B {\bf 529}, 424 (1998);
 W.~Bernreuther, A.~Brandenburg, Z.~G.~Si and P.~Uwer,
  Nucl.\ Phys.\ B {\bf 690}, 81 (2004);
 for a review see \eg:
 M.~Beneke {\it et al.},
  hep-ph/0003033.
 
\bibitem{resummation}
 J.~C.~Collins and D.~E.~Soper,
  Nucl.\ Phys.\ B {\bf 193}, 381 (1981)
  [Erratum-ibid.\ B {\bf 213}, 545 (1983)];
 J.~C.~Collins and D.~E.~Soper,
  Nucl.\ Phys.\ B {\bf 197}, 446 (1982);
 J.~C.~Collins, D.~E.~Soper and G.~Sterman,
  Nucl.\ Phys.\ B {\bf 250}, 199 (1985);
 C.~T.~H.~Davies, B.~R.~Webber and W.~J.~Stirling,
  Nucl.\ Phys.\ B {\bf 256}, 413 (1985);
 G.~Bozzi, S.~Catani, D.~de Florian and M.~Grazzini,
  hep-ph/0508068.

\bibitem{matching}
 S.~Catani, F.~Krauss, R.~Kuhn and B.~R.~Webber,
  JHEP {\bf 0111}, 063 (2001);
  L.~L\"onnblad,
  JHEP {\bf 0205}, 046 (2002);
 P.~Nason,
  JHEP {\bf 0411}, 040 (2004);
 Z.~Nagy and D.~E.~Soper,
  hep-ph/0503053.

\bibitem{mrw}
 S.~Mrenna and P.~Richardson,
  JHEP {\bf 0405}, 040 (2004).
 
\bibitem{mcnlo}
  S.~Frixione and B.~R.~Webber,
  JHEP {\bf 0206}, 029 (2002);
  S.~Frixione, P.~Nason and B.~R.~Webber,
  JHEP {\bf 0308}, 007 (2003).

\bibitem{sherpa}
 S.~H\"oche, F.~Krauss, A.~Sch\"alicke, S.~Schumann and J.~C.~Winter,
  JHEP {\bf 0402}, 056 (2004);

\bibitem{wplusjets} 
 see \eg:
 W.~T.~Giele, T.~Matsuura, M.~H.~Seymour and B.~R.~Webber,
  FERMILAB-CONF-90-228-T;
 M.~H.~Seymour,
  Comput.\ Phys.\ Commun.\  {\bf 90}, 95 (1995);
 J.~Huston, I.~Puljak, T.~Sj\"ostrand and E.~Thom\'e,
  hep-ph/0401145 (in hep-ph/0403100);
 F.~Krauss, A.~Sch\"alicke, S.~Schumann and G.~Soff,
  hep-ph/0503280;
 N.~Lavesson and L.~L\"onnblad,
  JHEP {\bf 0507}, 054 (2005).
 
\bibitem{q2merge}
 G.~Miu and T.~Sj\"ostrand,
  Phys.\ Lett.\ B {\bf 449}, 313 (1999).
\end{thebibliography}
\end{document}